\newcommand{\rem}[1]{}
\documentclass{amsart}
\usepackage{amsfonts,amssymb,amsmath,amsthm,mathrsfs}
\usepackage{url}
\usepackage[dvips]{epsfig}
\newtheorem{thrm}{Theorem}[section]

\newtheorem{prop}[thrm]{Proposition}

\newtheorem{remark}[thrm]{Remark}
\theoremstyle{definition}

\begin{document}
\author[C.~A.~Mantica and L.~G.~Molinari]
{Carlo~Alberto~Mantica and Luca~Guido~Molinari}
\address{C.~A.~Mantica: 
I.I.S. Lagrange, Via L. Modignani 65, 
20161, Milano, Italy -- L.~G.~Molinari (corresponding author): Physics Department,
Universit\`a degli Studi di Milano and I.N.F.N. sez. Milano,
Via Celoria 16, 20133 Milano, Italy.}
\email{carloalberto.mantica@libero.it, luca.molinari@unimi.it}
\title[Twisted manifolds]
{Twisted Lorentzian manifolds\\ a characterization with torse-forming \\ time-like unit vectors}
\begin{abstract} 
Robertson-Walker and Generalized Robertson-Walker spacetimes may be characterized by the
existence of a time-like unit torse-forming vector field, with other constrains. We show that Twisted manifolds
may still be characterized by the existence of such (unique) vector field, with no other constrain.
Twisted manifolds generalize RW and GRW spacetimes by admitting a scale function that depends both on time and space.
We obtain the Ricci tensor, corresponding to the stress-energy tensor of an imperfect fluid. 
\end{abstract}
\date{4 march 2017}
\maketitle
\section{\bf Introduction}
There is a hierarchy of Lorentzian manifolds $\mathscr L_n$ that, in privileged coordinates, gain the metric structure
\begin{align}
ds^2 = -dt^2 + f^2 g^*_{\mu\nu} (\vec x) dx^\mu dx^\nu \label{eq_twisted}
\end{align}
where $f>0$ is the scale factor and $g^*_{\mu\nu}$ is the metric tensor of a Riemannian sub-manifold $M^*$ of dimension $n-1$. The
scale factor depends on time, for otherwise the manifold is a product of disjoint manifolds.

1) The first  in the hierarchy are the {\sf Robertson-Walker (RW) spacetimes}, homogeneous and isotropic in space. They first appeared in the works by Friedmann (1922)
as solutions of Einstein's field equations, and were later derived on the basis of symmetries by Robertson (1935) and Walker (1936).
They model the standard large-scale cosmology \cite{Weinberg}. The scale function depends only on time and $M^*$ is a constant curvature Riemannian manifold, i.e. its Riemann tensor is:
$$R^*_{\mu\nu\rho\sigma} = \frac{R^*}{(n-1)(n-2)} (g^*_{\mu\rho} g^*_{\nu\sigma}-g^*_{\mu\sigma}g^*_{\nu\rho} ).  $$  
A covariant expression of the Riemann tensor $R_{jklm}$ of $\mathscr L_n$ is given in \cite{ManMolJMP};
the Weyl tensor $C_{jklm}$ is zero. The Ricci tensor $R_{ik}=R_{ijk}{}^j$ has the ``perfect fluid" form $Au_iu_j +Bg_{ij}$, with scalar fields $A$, $B$ and a unit time-like vector field, $u^ku_k=-1$, which is torse-forming:
\begin{align}
\nabla_i u_j = \varphi (u_iu_j + g_{ij}) \label{eq_tsv} 
\end{align}
and is an eigenvector of the Ricci tensor.

2) Next come the {\sf Generalized RW (GRW) spacetimes}, where space homogeneity is relaxed by allowing $M^*$ to be any Riemannian manifold, 
with the scale factor still being a function of time only (so $\mathscr L_n$ is a warped manifold). They were introduced in 1995 by Al{\`\i}as, Romero and S\'anchez \cite{AliRomSan95_a} 
and their geometric properties have been intensely studied \cite{survey}. RW spacetimes are the subclass with zero Weyl tensor \cite{Brozos}. The slices $\{t\}\times M^*$ are totally umbilical, with the average of the $n-1$ principal curvatures being constant on the hypersurface, $H=\dot f/f$
(this is Hubble's parameter in standard cosmology). \\
B-Y Chen gave in 2014 a very simple characterization \cite{Chen14}: a Lorentzian manifold is a GRW if and only if there exists a time-like concircular vector field: $X^kX_k <0$ and $\nabla_i X_j = \rho g_{ij}$. The time-like unit vector field $u_k = X_k/\sqrt{-X^2}$ is torse-forming, eq.\eqref{eq_tsv}, and is an eigenvector of the Ricci tensor. If $\xi $ is the eigenvalue and $R$ is the scalar curvature, the general form of the Ricci tensor in GRW spacetimes is \cite{ManMolJMP}:
\begin{align}
 R_{ij} = -(n-2) u^ru^s C_{rijs} - \frac{n\xi -R}{n-1} u_iu_j + \frac{R-\xi}{n-1} g_{ij} 
 \end{align}  
In the same paper the following equivalence was proven:  $\nabla^m C_{jklm}=0$ if and only if $u^m C_{jklm}=0$, which is the condition for the Ricci tensor to simplify into the perfect fluid form.

3) A further generalization are the {\sf Twisted spacetimes}, where the scale factor $f$ depends on both time and position and $M^*$ is a Riemannian manifold. They were introduced by B-Y Chen in 1979 (\cite{Chen1979}, definition 8.1) as the natural generalization of warped manifolds that avoids the
constancy of mean curvature of slices. Very recently Chen gave a simple characterization 
(\cite{Chen17}, theorem 3.2): {\em A Lorentzian manifold $\mathscr L_n$ admits a time-like torqued vector field $\tau $, i.e. 
$$\tau^i\tau_i<0, \qquad  \nabla_i \tau_j = \varphi g_{ij} + \alpha_i \tau_j, \quad \alpha_i\tau^i =0 $$
if and only if it is locally a twisted product $I\times_f M^*$, where $I$ is an open interval, $M^*$ is a
Riemannian (n--1)-manifold. } 

In this paper we characterize twisted spaces again through the existence of a time-like unit and torse-forming vector field, 
eq.\eqref{eq_tsv}, which thus becomes a unifying property for the metrics \eqref{eq_twisted}. For twisted spaces this suffices, for GRW it must be  
an eigenvector of the Ricci tensor, for RW one must also require $C_{jklm}=0$.

We show that the torse-forming time-like unit vector field is unique and Weyl compatible. We give the general form of the Ricci tensor in Lorentzian twisted spaces, and comment on its relation with the stress-energy tensor of imperfect fluids.

\section{\bf A characterization by torse-forming vectors}
The proof for characterizing twisted spaces by the existence of a time-like unit torse-forming vector field
relies on the following assertion, excerpted from a theorem by Ponge and Reckziegel \cite{PonRec93}: {\em
Let $(M,g)$ be a pseudo-Riemannian space with $M=B\times F$ and assume that the canonical foliations $L_B$ and $L_F$ intersect perpendicularly everywhere. Then $g$ is the metric tensor of a twisted product $B \times_b F$ if and only if $L_B$ is a totally geodesic foliation and $L_F$ is a totally umbilic foliation}.

Given the time-like unit vector $u_i$, we denote $h_{ij}=u_iu_j+ g_{ij}$ the projection matrix on the perpendicular direction. The torse-forming property is $\nabla_i u_j =\varphi h_{ij}$.

\begin{thrm} 
A Lorentzian manifold $\mathscr L_n$ is twisted  if and only if it admits a torse-forming time-like unit vector field. 
\begin{proof}
Let $\mathscr L_n$ be a Lorentzian twisted manifold, then there is a frame where the metric has the form \eqref{eq_twisted}.  The Christoffel symbols are listed 
in the Appendix.
The time-like unit vector field with components $u_0=-1$, $u_\mu =0$ identically solves the equation $\nabla_i u_j = \varphi (u_i u_j + g_{ij})$ with $\varphi = \dot f/f $ (the non-trivial
equations are $0=\varphi (u_0^2 + g_{00})$ and $ -\Gamma_{\mu\nu}^0 u_0 = \varphi f^2 g^*_{\mu\nu}$, which yields $\varphi $). 

The other way, suppose that a Lorentzian manifold is endowed with a vector field $\nabla_i u_j =\varphi h_{ij}$,  $u^k u_k=-1$. It is $u^k\nabla_k u_j=0$ ($u$ is geodesic) and $\nabla_i u_j = \nabla_j u_i$ ($u$ is closed). Being $u_i = \nabla_i \vartheta $, it is 
the unit normal vector field for the surfaces $\vartheta = const$. 
Any vector $V\in T_P(\mathscr L_n)$ is decomposable into a normal component and a component tangent to the hypersurface: $V^i=(u^kV_k) u^i + \tilde V^i$, $\tilde V^i=h^i{}_j V^j$. 
The induced metric is $\tilde g (\tilde V, \tilde V') = g(hV, hV')= h_{km} \tilde V^k \tilde V^{\prime m}$.\\
The second form of the hypersurface is defined by the relation $(hV)^k\nabla_k u_j =\Omega_{ij} (hV)^i $. Since the normal vector is torse-forming, then 
$\varphi (hV)^k h_{kj} = \Omega_{ij} (hV)^i$ i.e. $\Omega_{ij} = \varphi h_{ij}$. The hypersurface is  totally umbilical.\\
Since the manifold decomposes into a totally geodesic foliation orthogonal to a totally umbilical foliation, according to the assertion by
Ponge and Reckziegel, the metric has the twisted form.
\end{proof}
\end{thrm}

\begin{remark}
The mean curvature of a slice $\{t\}\times M^*$ is not uniform on the slice: $H(t,\vec x)=\varphi (t,\vec x)$.
The covariant expression is $\varphi =  u^k\nabla_k \log f $.
 \end{remark}

\begin{prop}
In a twisted manifold with $\dot f \neq 0$, the torse-forming time-like unit vector field is unique (up to a sign).
\begin{proof}
Suppose that, besides the vector in the theorem, there is another time-like unit torse-forming vector field, $\nabla_iv_j = \lambda (v_i v_j+ g_{ij})$.
In the frame \eqref{eq_twisted}  the condition $v^kv_k=-1$ is $(v_0)^2- \frac{1}{f^2} g^{*\mu\nu} v_\mu v_\nu =1$.   Put $v_0 =  \cosh a $ and $v_\mu = V_\mu \, \sinh a $, 
with $a (t,\vec x)\neq 0$ and $g^{*\mu\nu} V_\mu V_\nu =  f^2$.
The equation $\nabla_0 v_0 = \lambda (v^2_0 - 1)$ gives  $\dot a \, \sinh a  = \lambda \sinh^2 a $. Then:
$\dot a =\lambda \,\sinh a  $. The equation $\nabla_0 v_\mu = \lambda (v_0 v_\mu )$ i.e.
\begin{align*}
\dot a \, \cosh a \, V_\mu + \sinh a \, \dot V_\mu + \frac{\dot f}{f} \sinh a \, V_\mu = \lambda \cosh a \sinh a V_\mu 
\end{align*}  
simplifies to: $\sinh a ( f\dot V_\mu + V_\mu \dot f )=0$. If $a\neq 0$ it is $fV_\mu$ independent of $t$. This result is compatible with
$g^{*\mu\nu} V_\mu V_\nu =  f^2$ only in the trivial case $\dot f =0$. 
\end{proof}
\end{prop}

\section{\bf The Ricci tensor}
Let us introduce the orthogonal decomposition $\nabla_i \varphi = v_i - u_i u^k \nabla_k\varphi $, where $v_i = h_i{}^k\nabla_k \varphi $. 
In the frame \eqref{eq_twisted} it is $v_0=0$, $v_\mu=\partial_\mu \varphi $.

\begin{prop}
The Ricci tensor on  a twisted Lorentzian space  has the form
\begin{align}
R_{kl} =  - ( n u_ku_l +g_{kl})  (u^r\nabla_r \varphi + \varphi^2)  +  \tfrac{R}{n-1} \, h_{kl}   \label{eq_Ricci}\\
 + (n-2) (u_kv_l + u_l v_k - u^ru^s C_{rkls}) \nonumber
\end{align}
where $R$ is the curvature scalar, $C_{jklm}$ is the Weyl tensor, $v_k = h_k{}^m \nabla_m\varphi $.
\begin{proof}
Let us evaluate  $\nabla_j (\nabla_k u_l) = \nabla_j ( \varphi h_{kl} )$ and subtract the expression with $j,k$ exchanged. We obtain:
\begin{align}
R_{jkl}{}^m u_m = h_{kl}\nabla_j\varphi - h_{jl} \nabla_k \varphi  + \varphi^2 (u_k g_{jl} -  u_j g_{kl}) \label{Riemann_u}
\end{align}
The contraction with $g^{kl}$ gives:
\begin{align}
R_j{}^m u_m = -(n-2)v_j + (n-1)(u^k\nabla_k \varphi + \varphi^2) u_j \label{eq_Ricciu}
\end{align}
The Weyl tensor, contracted with $u$ is:
\begin{align}
C_{jkl}{}^m u_m = R_{jkl}{}^m u_m + \tfrac{1}{n-2} [ u_j R_{kl} - u_k R_{jl}+g_{kl}R_{jm}u^m -g_{jl}R_{km}u^m ] \label{eq_Weylu} \\
- \tfrac{1}{(n-1)(n-2)}R(u_j g_{kl}-u_k g_{jl}). \nonumber
\end{align}
Another contraction singles out $R_{kl}$, and the result is obtained. 
\end{proof}
\end{prop}

In the twisted frame \eqref{eq_twisted} the components $R_{00}$ and $R_{\mu0}$ listed in the Appendix are recovered. The comparison for the components $R_{\mu\nu}$ yields the identification:
\begin{align}
- (n-2) C_{0\mu\nu 0} = R^*_{\mu\nu} - R^*\frac{g^*_{\mu\nu}}{n-1} + 2(n-3) \left[ \frac{f_\mu f_\nu}{f^2} - \frac{g^*_{\mu\nu}}{n-1}  \frac{f_\sigma f^\sigma}{f^2}\right ] \\
-(n-3)\left [ \frac{\nabla_\mu^* f_\nu}{f} - \frac{g^*_{\mu\nu}}{n-1} \frac{\nabla_\sigma^* f^\sigma }{f}\right ] \nonumber
\end{align}

Multiplication of \eqref{eq_Weylu} and \eqref{Riemann_u} by $u_i$ and summation on cyclic permutation of indices $ijk$, after
some computation and cancellations, shows that the tensor $u_i u_m$ is Weyl compatible 
(see \cite{ManMol_Weyl}):
\begin{align} \label{Weylcomp}
u_i u^m C_{jklm} + u_j u^m C_{kilm} + u_k u^m C_{ijlm} =0 
\end{align}
It follows that:\\
1) $ u^m C_{jklm} =   u_k (u^iu^m C_{ijlm}) -u_j (u^i u^m C_{iklm})$. Therefore $u^m C_{jklm}=0$ if and only if $u^i u^m C_{iklm}=0$.\\
2) According to the Bel-Debever criterion, a twisted space-time is purely electric (see  \cite{Hervik2013} proposition 3.17).\\

\begin{prop}
A twisted manifold is a GRW if and only if: $h_i{}^k \nabla_k \varphi =0$
\begin{proof}
In Proposition 3.7 of \cite{survey} we proved that a Lorentzian manifold is a GRW spacetime if and only if it admits a torse-forming time-like
unit vector that is also eigenvector of the Ricci tensor. By eq. \eqref{eq_Ricciu} this is true if and only if $v_i=0$. 
\end{proof}
\end{prop}

We end with some remarks on imperfect fluids.
The form \eqref{eq_Ricci} of the Ricci tensor makes twisted spaces solutions of the Einstein's field equations for 
imperfect fluids. The Einstein's equations, $R_{ij} -\frac{1}{2}R g_{ij} = 8\pi T_{ij}$, give the energy-stress tensor
\begin{gather}
 T_{ij}=(p+\mu) u_i u_j + p g_{ij} + (q_iu_j + u_i q_j) + P_{ij}   \label{eq_energystress}   \\
 \mu = -\frac{n-1}{8\pi}(u^r\nabla_r\varphi +\varphi^2)+\frac{R}{16\pi}, \\
 p= -\frac{1}{8\pi} (u^r\nabla_r\varphi +\varphi^2) - \frac{R}{8\pi} \frac{n-3}{2(n-1)},  \\
 q_j = \frac{n-2}{8\pi} v_j , \quad P_{ij} = -\frac{n-2}{8\pi} u^ru^s C_{rijs}.
\end{gather} 
Eq.\eqref{eq_energystress} is the form that describes an imperfect fluid with velocity field $u^i$, anisotropic stress tensor $P_{ij}$ (with the properties $P_{ij}u^j=0$, $P^j_j=0$), energy flux $q_j$ (with $q_ju^j=0$), effective pressure $p$ and energy density $\mu $ \cite{Maartens}. \\
Since $u_i$ is torse-forming, the acceleration, the shear and the vorticity tensors are zero. Such spacetimes have been investigated
in detail by Coley and McManus \cite{ColeyMcManus}.

Besides $u_i$, there is another relevant time-like unit velocity field, which is an eigenvector of $T_{ij}$ and then of $R_{ij}$:  
$ R_{ij} w^j = W w_i $,  $w^iw_i=-1$. \\
The simplest situation for evaluating it in a twisted space, is to require 
$u^ru^sC_{rkls} v^k= C v_l$ for some scalar field $C$. Then $w^j$ can be obtained as a linear combination 
of $u_j$ and $q_j$.


\section*{\bf Appendix}
\centerline{( $i,j,k,... =0,1,...,n-1$ ; \quad $\mu,\nu,\rho,... = 1,2, ... , n-1$)}

{\quad}\\ \noindent
{\sf Christoffel symbols}: $\Gamma_{ij}^k = \Gamma_{ji}^k = \tfrac{1}{2}g^{km} (\partial_i g_{jm} +\partial_j g_{im} -\partial_m g_{ij})$. 
\begin{gather}
\Gamma_{i,0}^0=0,\quad  \Gamma_{0,0}^k=0, \quad \Gamma^\rho_{\mu,0} = (\dot f/f ) \delta^\rho_\mu, \quad 
\Gamma^0_{\mu,\nu} = f\dot f g^*_{\mu\nu}, \\
\Gamma^\rho_{\mu,\nu} = \Gamma^{*\rho}_{\mu,\nu} + (f_\nu/f)  \delta^\rho_\mu + (f_\mu/f)  \delta^\rho_\nu - (f^\rho/f)   g^*_{\mu\nu} 
\end{gather}
where $\dot f =\partial_t f$, $f_\mu = \partial_\mu f$ and $f^\mu = g^{*\mu\nu} f_\nu $.\\

\noindent
{\sf Riemann tensor}: $R_{jkl}{}^m = -\partial_j \Gamma^m_{k,l} + \partial_k \Gamma^m_{j,l} + \Gamma_{j,l}^p\Gamma^m_{kp} - \Gamma_{k,l}^p
\Gamma_{jp}^m $
\begin{gather}
R_{\mu 0\rho}{}^0 = (f\ddot f) g^*_{\mu\rho} \\
 R_{\mu\nu\rho}{}^0 = g^*_{\mu\rho} (f \partial_\nu \dot f - \dot f f_\nu) - g^*_{\nu\rho} (f \partial_\mu \dot f - \dot f f_\mu ) 
\end{gather}
\begin{align}
R_{\mu\nu\rho}{}^\sigma =& \, R^*_{\mu\nu\rho}{}^\sigma + ({\dot f}^2 - \frac{f^\lambda f_\lambda}{f^2} ) (g^*_{\mu\rho}\delta^\sigma_\nu - g^*_{\nu\rho}\delta^\sigma_\mu)\\
&+\frac{2}{f^2} ( f^\sigma f_\nu g^*_{\mu\rho} - f^\sigma f_\mu g^*_{\nu\rho} + f_\mu f_\rho \delta^\sigma_\nu - 
f_\nu f_\rho \delta^\sigma_\mu )  \nonumber \\
&+\frac{1}{f} \left[ \nabla^*_\mu (f^\sigma g^*_{\nu\rho} - f_\rho  \delta^\sigma_\nu)  - \nabla^*_\nu ( f^\sigma g^*_{\mu\rho} - f_\rho  \delta^\sigma_\mu)
\right ] \nonumber
\end{align}

\noindent
{\sf Ricci tensor}: $R_{jl} = R_{jkl}{}^k $
\begin{align}
R_{00}  =& -(n-1) (\ddot f / f) \\
R_{\mu 0} =& -(n-2)\partial_\mu (\dot f / f) \\
R_{\mu\nu} =& R^*_{\mu\nu} + g^*_{\mu\nu} [(n-2){\dot f}^2 + f\ddot f] +2(n-3)\frac{f_\mu f_\nu}{f^2} -(n-4) \frac{f^\sigma f_\sigma }{f^2} g^*_{\mu\nu}\\
&-(n-3) \frac{1}{f}\nabla^*_\mu f_\nu  - \frac{1}{f} g^*_{\mu\nu} \nabla^*_\sigma f^\sigma   \nonumber
\end{align} 

\noindent
{\sf Curvature scalar}: $R=R^k{}_k $
\begin{align}
R=\frac{R^*}{f^2} +(n-1) \left [(n-2)\frac{{\dot f}^2}{f^2} + 2\frac{\ddot f}{f}\right ] - (n-2)(n-5)\frac{f^\sigma f_\sigma }{f^4} 
-2(n-2)\frac{\nabla^*_\sigma f^\sigma  }{f^3} 
\end{align}

\end{document}